# Mirror enhanced directional out-coupling of SERS by remote excitation of a nanowire-nanoparticle cavity

*Sunny Tiwari[a]\*, Adarsh B. Vasista[b], Diptabrata Paul[a], and G.V. Pavan Kumar[a]*

[a] Sunny Tiwari, Diptabrata Paul and G.V. Pavan Kumar
Department of Physics, Indian Institute of Science Education and Research, Pune-411008, India

\*E-mail: sunny.tiwari@students.iiserpune.ac.in

[b] Adarsh B. Vasista
Department of Physics and Astronomy, University of Exeter, EX44QL, United Kingdom

We report on the experimental observation of mirror enhanced directional surface enhanced Raman scattering (SERS) from a self-assembled monolayer of molecules coupled to a nanowire-nanoparticle (NW-NP) junction on a mirror in remote excitation configuration. Placing NW-NP junction on a metallic mirror generates multiple gap plasmon modes which have unique momentum space scattering signatures. We perform Fourier plane imaging of SERS from NW-NP on a mirror to understand the effect of multiple hotspots on molecular emission. We systematically study the effect of ground plane on the directionality of emission from NW-NP junction and show that the presence of a mirror drastically reduces angular spread of emission. The effect of multiple hotspots in the geometry on directionality of molecular emission is studied using 3D numerical simulations. The results presented here will have implications in understanding plasmon hybridization in the momentum space and its effects on molecular emission.

Controlling photo-physics of molecules has far reaching implications in understanding many aspects of quantum physics[1] and in applications such as biomedical imaging,[2] designing molecular antennas,[3-5] and catalysis[6,7]. A simple way to achieve this control is to couple molecules to confined optical fields like plasmonic cavities,[8-10] whispering gallery microcavities[11] and Fabry-Pérot cavities.[12]

When a molecular dipole is placed inside an optical cavity, its emission characteristics such as rate of spontaneous emission,[9,13] polarization signatures,[14,15] and direction of emission[4,14] can be influenced. The ability to engineer plasmon-matter interactions has been extensively utilized to design and develop optical antennas to direct optical emission from molecules.[1,5,16] An important aspect of optical antenna design is to achieve low angular spread without compromising the enhancement of light-molecule interactions. To this end, a variety of antennas have been studied to influence secondary emission from molecules and quantum dots.[17-19]

Of interest to this study is the emission of secondary photons through Raman scattering.[20,21] Thanks to the development in nanoscale fabrication and synthesis, routing Raman scattered light from molecules by designing plasmonic geometries has gained prominence.[22,23] Most of the studies in the context of Raman optical antennas use lithographically fabricated structures and arbitrarily couple molecules to 'top-down" nanostructures.[24,25] A relatively less explored approach is to create Raman optical antennas by self-assembling chemically prepared colloidal nanostructures which are pre-coated with molecules. Molecule coated particle can be placed near a plasmonic structure to utilize the localized electric field provided by the cavity formed between the particle and nanostructures for enhanced spectroscopies. A unique nanostructure, in this regard is a one dimensional plasmonic nanowire. Coupling nanoparticle with a plasmonic nanowire offers the possibility of remotely exciting the cavity using nanowire as a waveguide which eliminates the damage caused to the molecules in the direct excitation of cavity.[26-28] Such geometries have been used in our past works to show enhanced directional spectroscopic signals.[25,29] Placing the nanowire-nanoparticle on a metallic mirror can generate particle-mirror particle cavity having ultra-small mode volumes.[30,31] Such types of cavities formed using mirror have been used for various studies such as strong coupling at room temperature,[32] large Purcell enhancement[33] and in tailoring the spectral signatures of two dimensional materials.[34] In addition to the enhancement, metal substrates direct majority of the emission to the collection objectives,[16-35] because of the absence of leakage radiation which occurs in using high refractive index substrates.[1]

Motivated by this, we modify the geometry of nanowire-nanoparticle junction used in our past studies, [25,29] by placing the junction on a gold mirror and with a superior control on the placement of molecules in the cavity. We study silver nanowire-gold nanoparticle (NW-NP) junction placed on a gold mirror in remote excitation configuration. The fields generated in the NW-NP and NP on mirror (NP-Mirror) cavities enhance the Raman scattering from the molecules coated on the nanoparticle and also reduces the angular spread of emission.

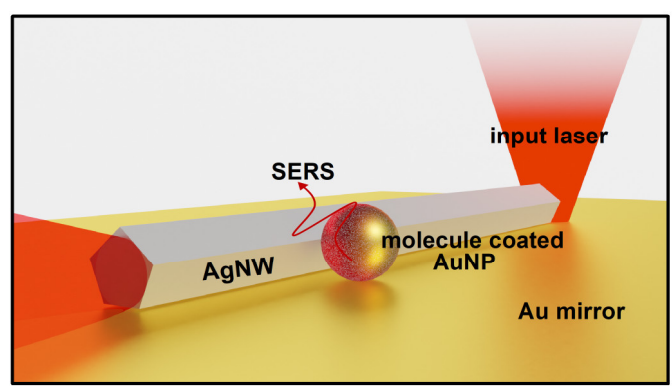

**Figure 1.** Schematic of the experimental configuration. A BPT coated AuNP was assembled near an AgNW placed on a gold mirror using self-assembly. One end of the AgNW was excited with a focused 633 nm laser. The SERS emission from the NW-NP junction was collected and its spectral and wavevector distribution were studied using spectroscopy and Fourier plane and energy-momentum imaging.

A schematic of the experimental configuration is shown in **figure 1**. Gold nanoparticles of size ~180 nm were coated with a monolayer of BPT molecules using self-assembly process. AgNWs of diameter ~350nm were prepared using polyol process[36] with polyvinylpyrrolidone (PVP) as a surfactant.[37] Gold mirrors of thickness 160 nm were prepared by thermally evaporating gold on a glass coverslip. A typical NW-NP junction was prepared through capillary force driven self-assembly,[38] by dropcasting silver nanowires on gold mirror, followed by drying and dropcasting of gold nanoparticles on top of it. One end of the AgNW was excited with a 633 nm laser using a 100x, 0.95 numerical aperture objective lens. AgNW surface plasmon polaritons (SPPs) get scattered by NW-NP junction and out-couples as free space photons and also excites the gap plasmons in the NW-NP cavity. By focusing light onto one end of the nanowire, we also excite SPPs on the metal film. These propagating plasmons on the metal film excite the gap mode between the nanoparticle and the mirror.[39] The gap plasmons generated at both the cavities enhance the Raman scattering signatures of the molecules coated on the NP and influence its far-field scattering signatures. Out-coupled free space emission from the junction was collected by the same objective lens by spatially filtering the region and was projected onto the Fourier plane to study the spectral and wavevector signatures (see supplementary information S1 and S2 for details on sample preparation and experimental setup respectively).

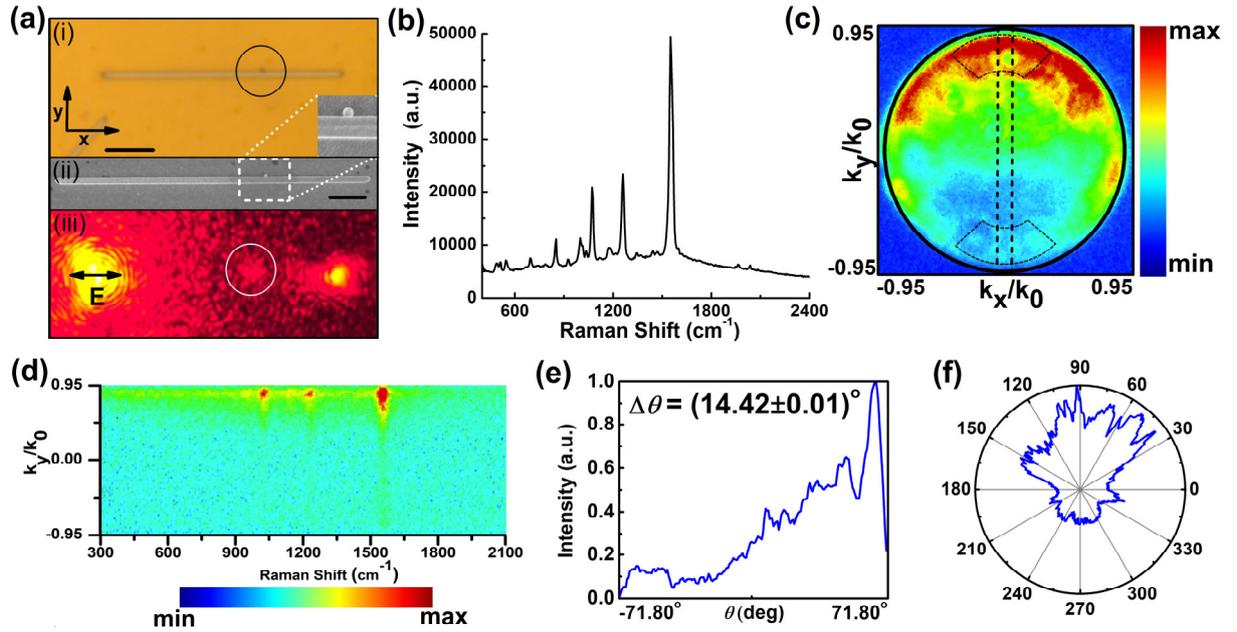

**Figure 2.** Remote excitation and Fourier plane imaging of SERS emission. (a) Optical imaging: (i) Bright field image of a ~180 nm AuNP coated with BPT molecules coupled to an AgNW placed on a gold mirror. Scale bar is 3μm. (ii) Scanning electron microscopy image of the same NW-NP junction. Scale bar is 3μm. (iii). Elastic scattering image upon excitation of one end of AgNW with incident light polarized along the AgNW. (b) SERS spectrum of remotely excited BPT molecules collected from the NW-NP on mirror junction. (c) Fourier plane image of SERS emission collected from the NW-NP junction. (d) Energy-momentum image captured after dispersing a small portion of Fourier plane image (shown in dotted black rectangle around $k_x/k_0$ = 0) shows the directional out-coupling of the Raman lines and inelastic background. (e) Intensity cross-cut along the $k_x/k_0$ = 0 line in the Fourier plane image. (f) The intensity profile of azimuthal angles ($\phi$) for $\theta$ corresponding to maximum intensity in the Fourier plane image (c).

**Figure 2a(i)** shows bright field image of a ∼180 nm NP coated with BPT molecules, assembled near an AgNW of thickness ∼350 nm placed on a 160 nm thick gold mirror. The black circle indicates the NW-NP junction. The corresponding scanning electron microscopy image of the junction is shown in figure 2a (ii). Figure 2.a (iii) shows the same AgNW when one end was excited with a focused 633 nm laser polarized along the axis of the nanowire. SPPs on AgNW and the metal film remotely excite the gap plasmons in the NW-NP and NP-Mirror cavities respectively. Intense electric field in the cavities due to these generated gap plasmons, results in the enhanced Raman scattering from the BPT molecules coated on the particle. We used a thick nanowire, diameter ∼350 nm, to get better waveguiding properties,[40,41] as we probed the NW-NP using remote excitation mechanism. The size of nanoparticle was chosen such that the localized plasmon resonance of the NP overlapped with the wavelength of the excitation to generate maximum response from the system. The out-coupled SERS emission from the junction (shown in a white circle in figure 2a (iii)) was collected. The remotely excited SERS spectrum of the BPT molecules from NW-NP junction on mirror is shown in Figure 2b. The sharp Raman lines are clearly visible with abroad inelastic background emission. Since the molecules are present only on the nanoparticle, the SERS emission originates only from the junction and not from the nanowire on the mirror cavity (see supporting information S3).

To study the wavevector distribution of remotely excited SERS emission, we performed Fourier plane imaging,[42,43] which quantifies the directionality of emission in terms of radial ($\theta$) and azimuthal angles ($\phi$). The Fourier plane image (figure 2.c) shows that the maximum SERS emission is biased towards higher $k_y/k_o$. The SERS emission from the NW-NP on mirror is more directional when the junction is excited remotely as compared to the direct excitation of the junction or only the NP-Mirror cavity (see supplementary information S4).

Along with the SERS emission from the BPT molecules, there is also an inelastic background emission from the PVP coating[37] on the nanowire which can also out-couple at higher angles (see supplementary information S5). To further confirm that the majority of emission at higher $+k_y/k_o$ angles is the SERS emission from the molecules we performed energy-momentum imaging[14,44] on the emission from the junction. A small portion of the Fourier plane image along $k_x/k_o = 0$ was projected onto the slit of the spectrometer and dispersed to get the image shown in figure 2.d. The energy- momentum image reveals that both SERS signal and the inelastic background from the junction out-couples at higher wavevectors.

To quantify the emission, we defined directionality (Dir), using the ratio of forward and backward intensity of emission in Fourier plane,[45] as

$$\text{Dir} = 10\log_{10}\frac{\iint_{(\theta m-\delta 1,\varphi m-\delta 2)}^{(\theta m+\delta 1,\varphi m+\delta 2)} I(\theta,\phi)\sin(\theta)d\theta\, d\phi}{\iint_{(\theta m-\delta 1,\phi m-\pi-\delta 2)}^{(\theta m+\delta 1,\phi m-\pi+\delta 2)} I(\theta,\phi)\sin(\theta)d\theta\, d\phi} \qquad (1)$$

Where $\theta_m$ and $\phi_m$ are the radial and azimuthal angles with maximum emission. $I(\theta, \phi)$ is the intensity in the Fourier plane image. For the black dotted region in the Fourier plane image, the calculated directionality is 8.0±0.2 db. See supplementary information S6 for the details in the variation of the calculated directionality with a change in $\delta_1$ and $\delta_2$. Intensity cross-cut along $k_x/k_0 = 0$ in the Fourier plane image (figure 2c) shows that the emission is sharp with a full width at half maxima (FWHM) of (14.42±0.01)° (figure 2e). The intensity profile of azimuthal angles ($\phi$) for $\theta$ corresponding to maximum intensity in the Fourier plane image shows that majority of emission is going along the NW-NP junction.

To understand the effect of the ground plane on SERS emission from NW-NP junction, we studied the wavevector of emission from NW-NP junction placed on a glass substrate. **Figure 3(a)** shows the bright field and transmission image of a BPT molecule coated ~180 nm AuNP assembled near an ~350 nm thick AgNW placed on a glass substrate. Upon excitation of one end using 633 nm laser with polarization along the AgNW axis, the AgNW plasmons out-couple from the AgNW end and NW-NP junction. The SERS emission from the junction was spatially filtered and was projected onto the Fourier plane. The Fourier plane image (Figure 3.b) shows that the emission is biased towards the higher $k_y/k_o$ values, but the angular spread in emission is very broad as compared to the spreading in the presence of gold substrate (see figure 2c).

We calculated the near-field electric field using finite element method with COMSOL Multiphysics as a solver to study the effect of different hotspots on emission wavevectors. Fourier plane images were then calculated by projecting the near-field to the far-field using reciprocity argument.[46] We place oscillating x, y and z oriented dipoles to mimic the molecular emission at the hotspots of the geometry. AgNW was modelled with a pentagonal cross-section with an edge to edge thickness of 350 nm and length of 5μm. AuNP of diameter 180 nm is placed at a distance of 5 nm from the AgNW. This 5 nm gap is to model the PVP coating on the AgNW and molecular coating on the AuNP. The refractive indices of the material were taken from reference.[47] Figure 3.c shows the calculated near-field electric field at the NW-NP junction placed on a glass substrate in remote excitation configuration. One end of the AgNW was excited using focused Gaussian laser of 633 nm. The field at the junction is only concentrated in the NW-NP cavity (shown as α), from where the SERS signal will originate. To study the effect of this cavity on the emission wavevector we placed oscillating x, y and z oriented dipoles at a wavelength of 703 nm in the NW-NP cavity and calculated Fourier plane image after incoherently adding the far-field radiation patterns from individual dipoles. The wavelength of the dipolar source was set at 703 nm because the BPT molecules have a prominent Raman mode at this wavelength. The calculated Fourier plane image corroborates the experimentally observed wavevector distribution of SERS. The emission is broad in terms of wavevectors and covers large range of angles. In the case of gold substrate, along with a high field at the NW-NP cavity, there is also a hotspot in the NP-Mirror cavity (shown as β) (Figure 3e). In this case, we placed oscillating x, y and z oriented dipoles at a wavelength of 703 nm in both NW-NP and NP-Mirror cavities and incoherently add the far-field radiation patterns from individual dipoles. The simulated Fourier plane image shows that the emission is biased towards NW-NP junction and is confined to narrow range of wavevectors. The results show the importance of metallic substrates for enhancement and in directing the SERS emission.

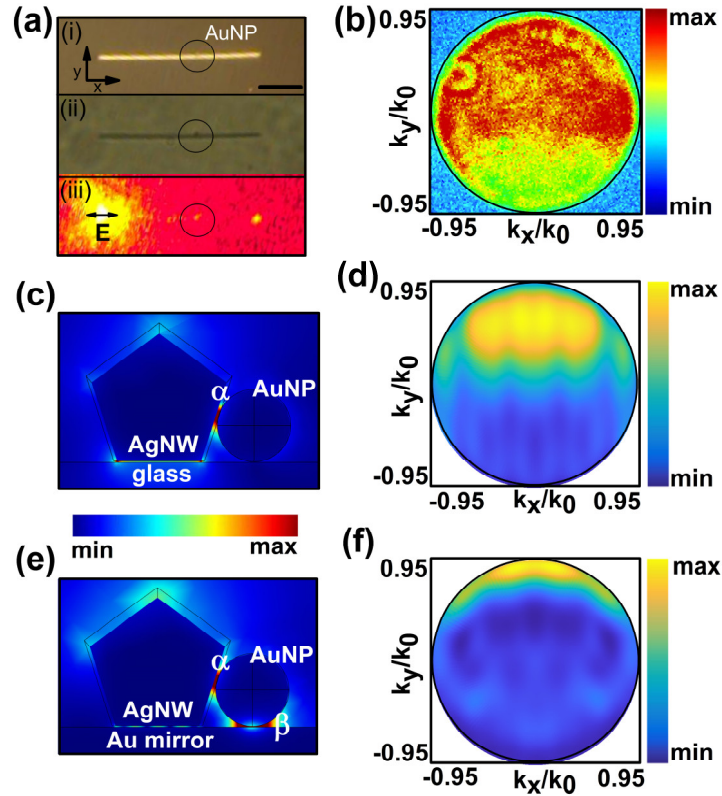

**Figure 3.** Role of Au mirror in influencing SERS wavevectors. (a) Optical imaging. (i) Bright field image of a BPT coated ~180 nm AuNP assembled near a ~350 nm thick AgNW placed on a glass substrate. Scale bar is 3μm. Transmission image of the same system. Elastic scattering image of the same system upon excitation of AgNW end using a 633 nm laser with polarization along the AgNW. (b) Fourier plane image of SERS emission collected from the NW-NP junction on glass substrate after rejecting the elastically scattered light. (c) Calculated near-field electric field at the NW-NP junction placed on a glass substrate in remote excitation configuration. (d) Calculated Fourier plane image after incoherently adding the far-field radiation patterns of x, y and z oriented dipoles placed in the gap between NW-NP (α). (e) Calculated near-field electric field at the NW-NP junction on mirror in remote excitation configuration. (f) Calculated Fourier plane image after incoherently adding the far-field radiation patterns of x, y and z oriented dipoles placed in the gap between NW-NP (α) and in the gap between NP-mirror (β). The wavelength of emission of dipolar source is 703 nm.

To study how AgNW influence the SERS emission wavevectors we study the change in the directionality of the emission when the distance between the AgNW and NP is varied. **Figure 4** shows the geometry of the system used in the calculations. Oscillating x, y, and z oriented dipoles at a wavelength of 703 nm ($\lambda$) are placed at the base of NP and the calculated near-field electric field was projected to the far-field. The calculated Fourier plane images show that the emission wavevectors are directional in nature when the distance between the nanowire and nanoparticle is between $\lambda/100$ to $\lambda/10$ as the nanowire reflects the emission along higher $k_y$ values. Without the presence of the nanowire, the emission pattern from an isolated nanoparticle on mirror cavity is isotropic in nature (see supplementary information S7). When the distance between the nanoparticle is changed to $\lambda/3$ the emission is no longer directional in nature but is still biased towards the higher k vectors. At a distance of $\lambda$, the influence of AgNW on the emission wavevectors is minimal and the emission is no longer biased or directional in any specific direction because at larger distances the reflection from the AgNW reduces. This shows that the nanowire not only facilitates the SERS enhancement but also directs emission in narrow range of angles.

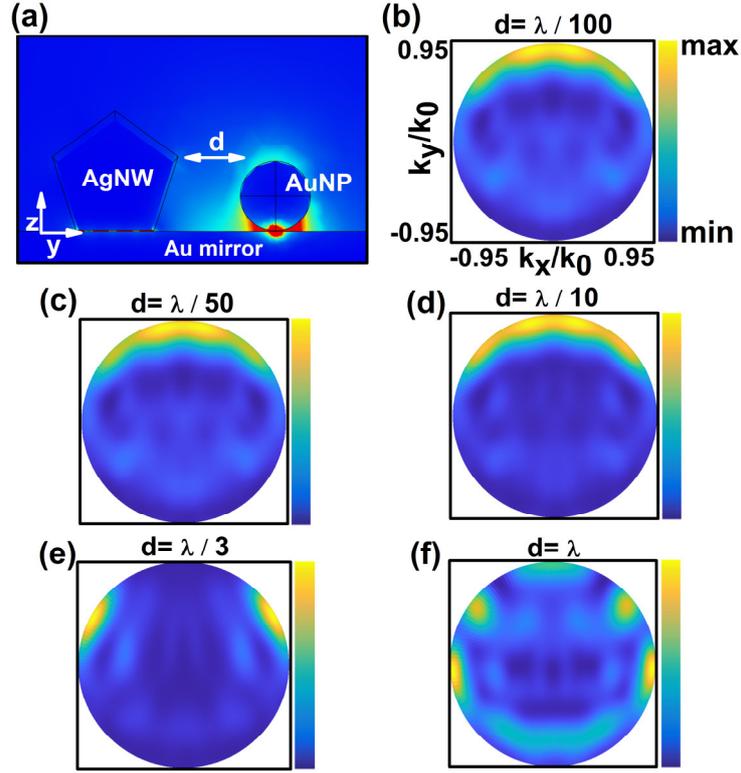

**Figure 4.** Variation in the directionality of emission with varying distance between nanoparticle and nanowire. (a) Geometry of NW-NP on mirror cavity with a distance 'd' between nanowire and nanoparticle. The nanowire (350 nm thickness) and nanoparticle (180 nm diameter) are placed on a gold mirror. (b-h) Calculated Fourier plane image when the distance between nanoparticle and nanowire is $\lambda/100$, $\lambda/50$, $\lambda/10$, $\lambda/3$ and $\lambda$ respectively. In each case, oscillating x, y and z oriented dipoles at wavelength of 703 nm ($\lambda$) are placed at the base of the nanoparticle and the Fourier plane image is calculated by incoherently adding the emission from individual dipoles.

To conclude, we have shown how unidirectional SERS emission can be achieved by NW-NP junction on mirror cavity in remote excitation configuration. The nanowire and gold mirror helps in providing enhancement and directing the SERS emission to a narrow range of wavevectors. Calculated forward-to-backward emission ratio for SERS emission is ~8 dB. Three-dimensional numerical calculations reveal the influence of electromagnetic hotspots generated in the geometry on the wavevectors of out-coupled SERS emission. We believe that the results shown in this letter will be extrapolated for studying strong interaction of molecules with extremely small cavities in remote excitation configurations. The NW-NP junction on mirror excited by nanowire plasmons will be a good testbed for remote detection of single molecules and studying quantum electrodynamics effects.

**Conflict of interest**

There are no conflicts to declare.

**Acknowledgements**

Authors thank Chetna Taneja, Vandana Sharma, Shailendra Chaubey, Utkarsh Khandelwal and Suryanarayan Banerjee for fruitful discussions. Authors thank Shailendra Chaubey for preparing the gold mirrors. This work was partially funded by Air Force Research Laboratory grant (FA2386-18-1-4118 R&D18IOA118), DST Energy Science grant (SR/NM/TP-13/2016) and Swarnajayanti fellowship grant (DST/SJF/PSA02/2017-18) to G V PK.

**References**

[1] L. Novotny, B. Hecht, *Principles of Nano-Optics*, Cambridge University Press, Cambridge, **2012**.
[2] X. Han, K. Xu, O. Taratula, K. Farsad, *Nanoscale*, **2019**, 11, 799.
[3] K. Hübner, M. Pilo-Pais, F. Selbach, T. Liedl, P. Tinnefeld, F. D. Stefani, G. P. Acuna, *Nano letters*, **2019**, 19, 6629.
[4] D. Wang, W. Zhu, M. D. Best, J. P. Camden, K. B. Crozier, *Nano letters*, **2013**, 13, 2194.
[5] L. Novotny, N. van Hulst, *Nature photonics*, **2011**, 5, 83.
[6] Z. Zhang, C. Zhang, H. Zheng, H. Xu, *Accounts of Chemical Research*, **2019**, 52, 2506.
[7] J. J. Baumberg, *Faraday Discussions*, **2019**, 214, 501.
[8] J. T. Hugall, A. Singh, N. F. van Hulst, *Acs Photonics*, **2018**, 5, 43.
[9] K. J. Russell, T.-L. Liu, S. Cui, E. L. Hu, *Nature Photonics*, **2012**, 6, 459.
[10] H. T. Miyazaki, Y. Kurokawa, *Physical review letters*, **2006**, 96, 097401.
[11] X. Jiang, A. J. Qavi, S. H. Huang, L. Yang, *arXiv preprint arXiv:1805.00062*, **2018**.
[12] J. P. Long, B. S. Simpkins, Acs *Photonics*, **2015**, 2, 130.
[13] S. Khatua, P. M. R. Paulo, H. Yuan, A. Gupta, P. Zijlstra, M. Orrit, Acs *Nano*, **2014**, 8, 4440.
[14] A. B. Vasista, H. Jog, T. Heilpern, M. E. Sykes, S. Tiwari, D. K. Sharma, S. K. Chaubey, G. P. Wiederrecht, S. K. Gray, G. V. P. Kumar, *Nano letters*, **2018**, 18, 650.
[15] A. B. Vasista, S. Tiwari, D. K. Sharma, S. K. Chaubey, G. V. P. Kumar, *Advanced Optical Materials*, **2018**, 6, 1801025.
[16] A. Ahmed, R. Gordon, *Nano letters*, **2011**, 11, 1800.
[17] Y. Wang, C. Li, G. Duan, L. Wang, L. Yu, *Advanced Optical Materials*, **2019**, 7, 1801362.
[18] G. Lu, J. Xu, T. Wen, W. Zhang, J. Zhao, A. Hu, G. Barbillon, Q. Gong, *Materials*, **2018**, 11, 1435.
[19] A. G. Curto, G. Volpe, T. H. Taminiau, M. P. Kreuzer, R. Quidant, N. F. van Hulst, *Science*, **2010**, 329, 930.
[20] E. Le Ru, P. Etchegoin, *Principles of Surface-Enhanced Raman Spectroscopy: and related plasmonic effects*, Elsevier, **2008**.
[21] S. Schlücker, *Angewandte Chemie International Edition*, **2014**, 53, 4756.

[22] J. Langer, D. Jimenez de Aberasturi, J. Aizpurua, R. A. Alvarez-Puebla, B. Auguié, J. J. Baumberg, G. C. Bazan, S. E. J. Bell, A. Boisen, A. G. Brolo, J. Choo, D. Cialla-May, V. Deckert, L. Fabris, K. Faulds, F. J. García de Abajo, R. Goodacre, D. Graham, A. J. Haes, C. L. Haynes, C. Huck, T. Itoh, M. Käll, J. Kneipp, N. A. Kotov, H. Kuang, E. C. Le Ru, H. K. Lee, J.-F. Li, X. Y. Ling, S. A. Maier, T. Mayerhöfer, M. Moskovits, K. Murakoshi, J.-M. Nam, S. Nie, Y. Ozaki, I. Pastoriza-Santos, J. Perez-Juste, J. Popp, A. Pucci, S. Reich, B. Ren, G. C. Schatz, T. Shegai, S. Schlücker, L.-L. Tay, K. G. Thomas, Z.-Q. Tian, R. P. Van Duyne, T. Vo-Dinh, Y. Wang, K. A. Willets, C. Xu, H. Xu, Y. Xu, Y. S. Yamamoto, B. Zhao, L. M. Liz-Marzán, *Acs nano*, **2020**, 14, 28.
[23] Q. Guo, T. Fu, J. Tang, D. Pan, S. Zhang, H. Xu, *Physical review letters*, **2019**, 123, 183903.
[24] A. R. L. Marshall, J. Stokes, F. N. Viscomi, J. E. Proctor, J. Gierschner, J.-S. G. Bouillard, A. M. Adawi, *Nanoscale*, **2017**, 9, 17415.
[25] A. Dasgupta, D. Singh, S. Tandon, R. P. N. Tripathi, G. V. P. Kumar, *Journal of Nanophotonics*, **2013**, 8, 1.
[26] S. Tiwari, C. Taneja, G. Kumar, *arXiv preprint arXiv:2008.01331*, **2020**.
[27] Y. Huang, Y. Fang, Z. Zhang, L. Zhu, M. Sun, *Light: Science & Applications*, **2014**, 3, e199.
[28] J. A. Hutchison, S. P. Centeno, H. Odaka, H. Fukumura, J. Hofkens, H. Uji-i, *Nano letters*, **2009**, 9, 995.
[29] A. B. Vasista, S. K. Chaubey, D. J. Gosztola, G. P. Wiederrecht, S. K. Gray, G. V. P. Kumar, *Advanced Optical Materials*, **2019**, 7, 1900304.
[30] F. Benz, M. K. Schmidt, A. Dreismann, R. Chikkaraddy, Y. Zhang, A. Demetriadou, C. Carnegie, H. Ohadi, B. de Nijs, R. Esteban, J. Aizpurua, J. J. Baumberg, *Science*, **2016**, 354, 726.
[31] J. Hao, T. Liu, Y. Huang, G. Chen, A. Liu, S. Wang, W. Wen, *The Journal of Physical Chemistry C*, **2015**, 119, 19376.
[32] R. Chikkaraddy, B. de Nijs, F. Benz, S. J. Barrow, O. A. Scherman, E. Rosta, A. Demetriadou, P. Fox, O. Hess, J. J. Baumberg, *Nature*, **2016**, 535, 127.
[33] N. Kongsuwan, A. Demetriadou, R. Chikkaraddy, F. Benz, V. A. Turek, U. F. Keyser, J. J. Baumberg, O. Hess, *Acs Photonics*, **2018**, 5, 186.
[34] J. Huang, G. M. Akselrod, T. Ming, J. Kong, M. H. Mikkelsen, *Acs Photonics*, **2018**, 5, 552.
[35] A. Ahmed, R. Gordon, *Nano letters*, **2012**, 12, 2625.
[36] Y. Sun, B. Mayers, T. Herricks, Y. Xia, *Nano letters*, **2003**, 3, 955.
[37] J. Hwang, Y. Shim, S.-M. Yoon, S. H. Lee, S.-H. Park, *RSC Advances*, **2016**, 6, 30972.
[38] P. Li, X. Yan, F. Zhou, X. Tang, L. Yang, J. Liu, *Journal of Materials Chemistry C*, **2017**, 5, 3229.
[39] Y. Li, H. Hu, W. Jiang, J. Shi, N. J. Halas, P. Nordlander, S. Zhang, H. Xu, *Nano letters*, **2020**, 20, 3499.
[40] H. Yang, M. Qiu, Q. Li, *Laser & Photonics Reviews*, **2016**, 10, 278.
[41] T. Shegai, V. D. Miljković, K. Bao, H. Xu, P. Nordlander, P. Johansson, M. Käll, *Nano letters*, **2011**, 11, 706.
[42] A. B. Vasista, D. K. Sharma, G. P. Kumar, *digital Encyclopedia of Applied Physics*, **2018**, 1.
[43] J. A. Kurvits, M. Jiang, R. Zia, *JOSA A*, **2015**, 32, 2082.
[44] C. M. Dodson, J. A. Kurvits, D. Li, R. Zia, *Optics letters*, **2014**, 39, 3927.
[45] D. Vercruysse, Y. Sonnefraud, N. Verellen, F. B. Fuchs, G. Di Martino, L. Lagae, V. V. Moshchalkov, S. A. Maier, P. Van Dorpe, *Nano letters*, **2013**, 13, 3843.
[46] J. Yang, J.-P. Hugonin, P. Lalanne, *Acs Photonics*, **2016**, 3, 395.
[47] P. B. Johnson, R. W. Christy, *Physical review B*, **1972**, 6, 4370.

# Supporting Information

**Table of Content:**

**S1: Sample preparation**

Single crystalline silver nanowires were prepared using polyol process[1]. The nanowires have a pentagonal cross-section with an average thickness (edge to edge) of ~350 nm. Gold film of thickness 160 nm was deposited on a 2 nm chromium coated glass coverslip using thermal vapor deposition. Gold nanoparticle were purchased from Sigma Aldrich.

To coat the molecules on the nanoparticles, the particles were first cleaned with acetone several times to remove the surfactant coating from the particles. The particles were then soaked in 1 mM solution of biphenyl-4-thiol in ethanol for 24 hours. BPT molecules and ethanol were purchased form Sigma Aldrich. Because of the presence of thiol bond, the molecules get attached to the gold surface strongly. The solution was then centrifuged and washed several times to remove the excess BPT molecules and then particles coated with the BPT molecules were transferred in ethanol solution.

NW-NP junction was prepared using self-assembly technique[2]. Silver nanowires dispersed in ethanol solution were dropcasted on a gold mirror and left to dry. After this, molecules coated gold nanoparticles were dropcasted on the gold substrate. Gold nanoparticles tend to sit near the nanowire forming a self-assembled junction as shown in figure S1.

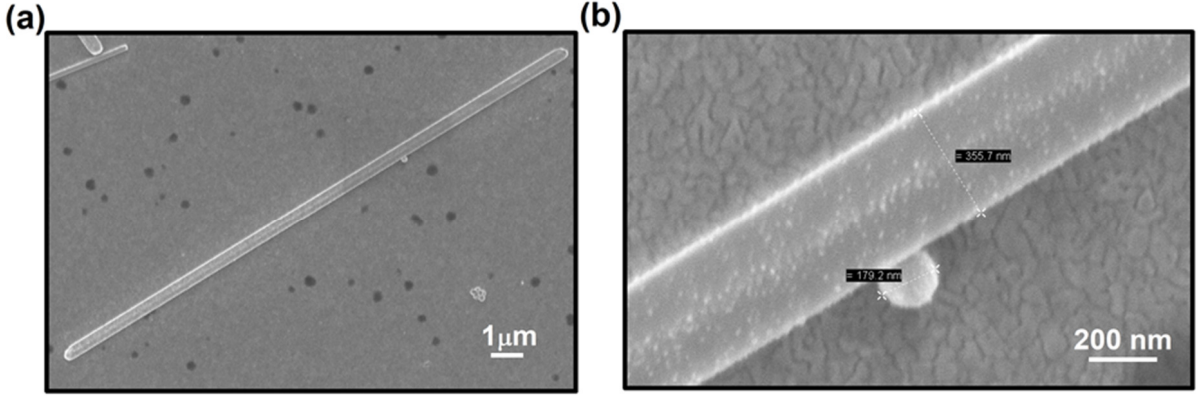

**Figure S1.** Scanning electron microscope image of a self-assembled NW-NP junction formed using a 356 nm thick silver nanowire and ~180 nm gold particle placed on a 160 nm thick gold mirror.

**S2: Experimental setup**

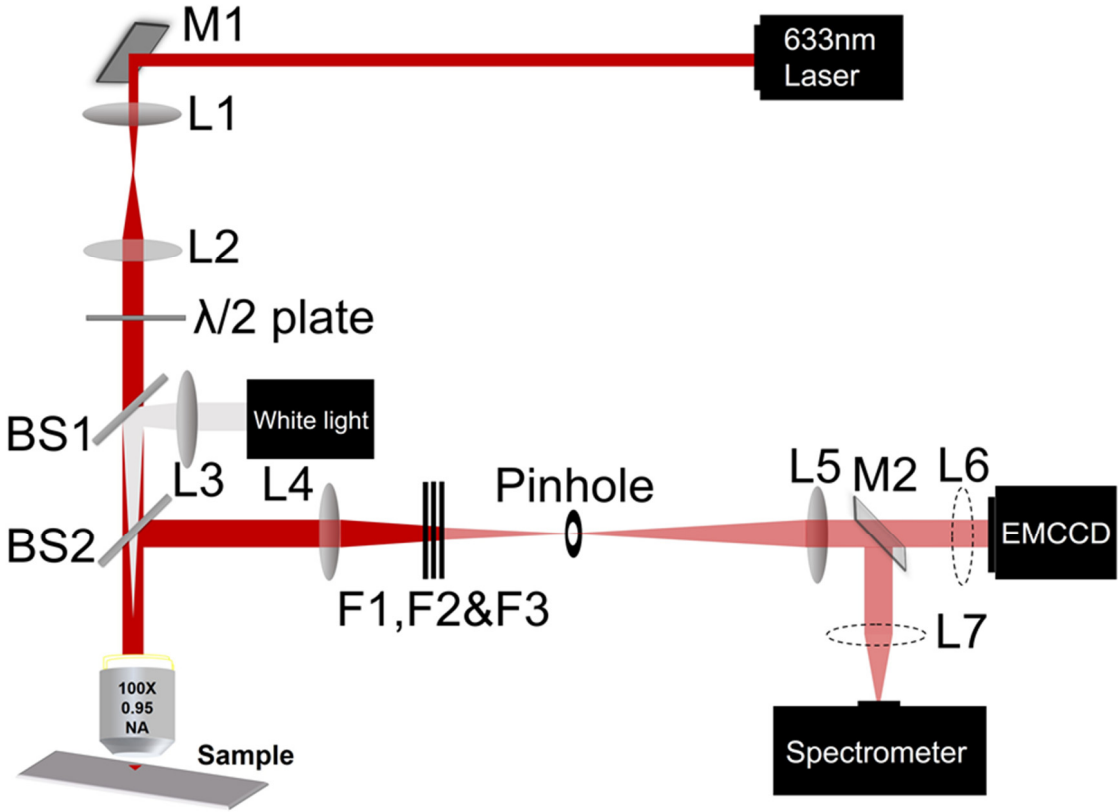

**Figure S2.** Schematic representation of the experimental setup.

The sample was excited using a high numerical aperture 100x, 0.95 NA objective lens. The backscattered light was collected using the same lens. The 633 nm laser light was expanded using a set of two lenses L1 and L2. M1 is a mirror. The polarization of the incoming laser was controlled by a λ/2 waveplate in the path. BS1 and BS2 are beam splitters to simultaneously excite the sample with laser and its visualization using white light. Lens L3 is used to loosely focus white light on the sample plane. F1, F2, and F3 are set of two edge filters and one notch filter to reject the elastically scattered light for SERS spectroscopy and Fourier plane[3] and energy-momentum imaging[4,5]. Lenses L4 and L5 are used to project the emission to the Fourier plane onto the spectrometer or EMCCD. M2 is a flip mirror, used to project the light on the spectrometer for spectroscopy and energy-momentum imaging. Lenses L6 and L7 are flip lenses used to switch from real plane to Fourier plane.

**S3: Spectrum from nanowire-nanoparticle on mirror cavity and from nanowire on mirror cavity**

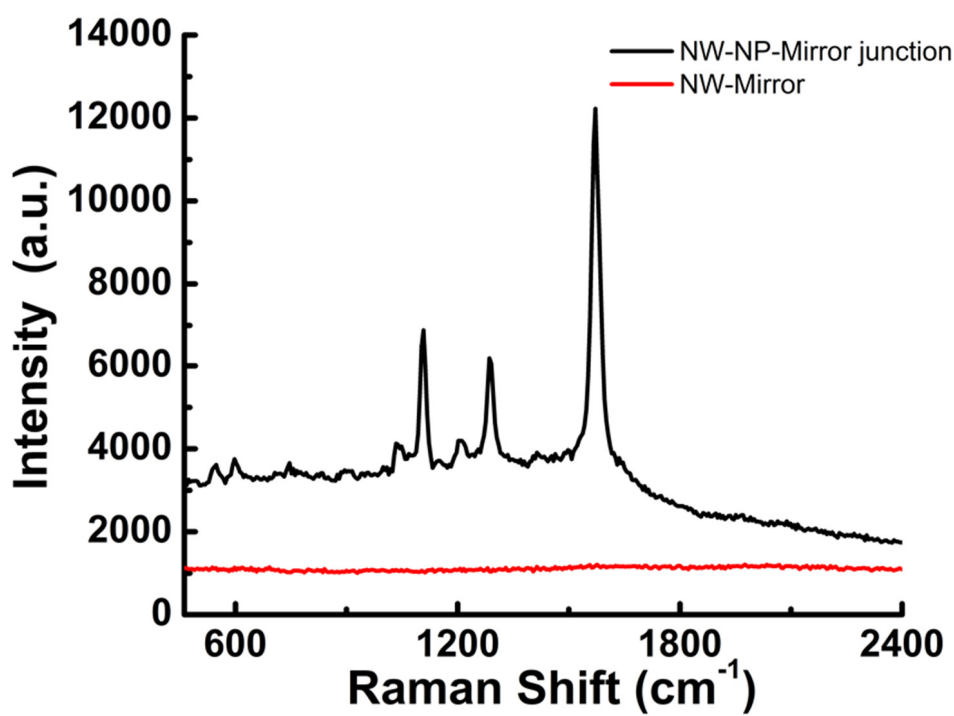

**Figure S3.** Spectra collected from the NW-NP on mirror cavity (black curve) and only from the nanowire on mirror cavity (red curve) with the same input intensity and exposure time. The spectra show that the emission originates only from the junction as the molecules are present only on the nanoparticle.

**S4: Fourier plane imaging of SERS emission from directly excited nanowire-nanoparticle on mirror and nanoparticle on mirror cavity**

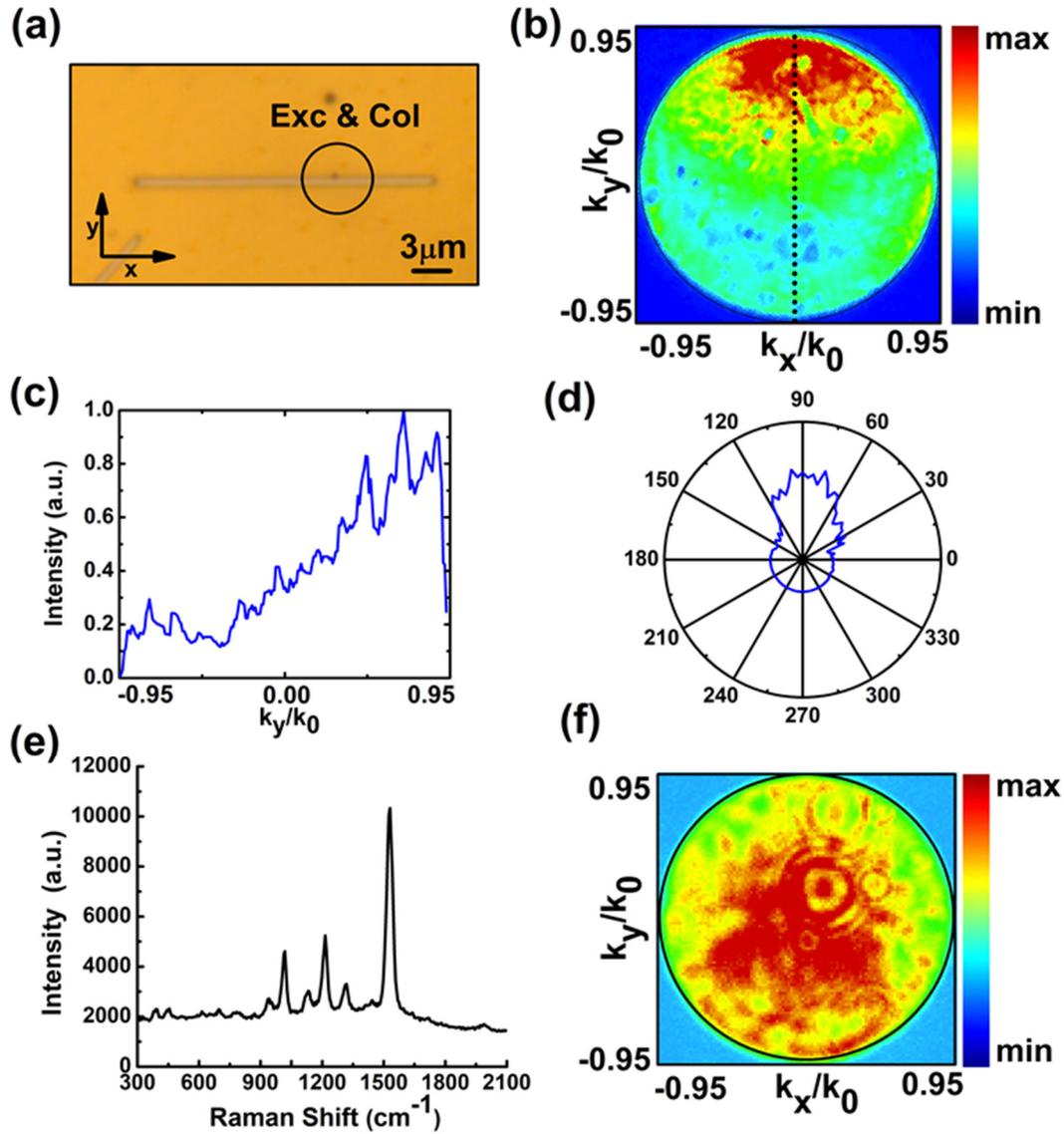

**Figure S4.** Fourier plane imaging of SERS emission collected from the direct excitation of NW-NP on mirror and NP-mirror cavities. (a) Bright field image of a ~180 nm diameter gold nanoparticle coated with BPT molecules attached to a silver nanowire (AgNW) placed on a gold mirror. The diameter of the nanowire is ~ 350 nm and the thickness of gold mirror is 160 nm. The junction was excited with a 633 nm laser polarized along the junction. (b) Fourier plane image of SERS emission from the junction after filtering out the elastically scattered photons. The emission is biased towards higher wavevectors but cover a broad range of wavevectors both in terms of radial and azimuthal angles. (c) Intensity cross-cut along the $k_x/k_0 = 0$ line in the Fourier image. (d) The intensity profile of azimuthal angles ($\phi$) for $\theta$ corresponding to maximum intensity in the Fourier plane image (c). The emission covers a broad range of azimuthal angles. (e) SERS spectrum from a BPT molecule coated ~180 nm diameter gold particle placed on a gold mirror. (f) Fourier plane image of SERS emission from the nanoparticle on mirror cavity. The emission is isotropic in nature and covers a large number of angles.

**S5: Inelastic emission from polyvinylpyrrolidone (PVP) coating on nanowire**

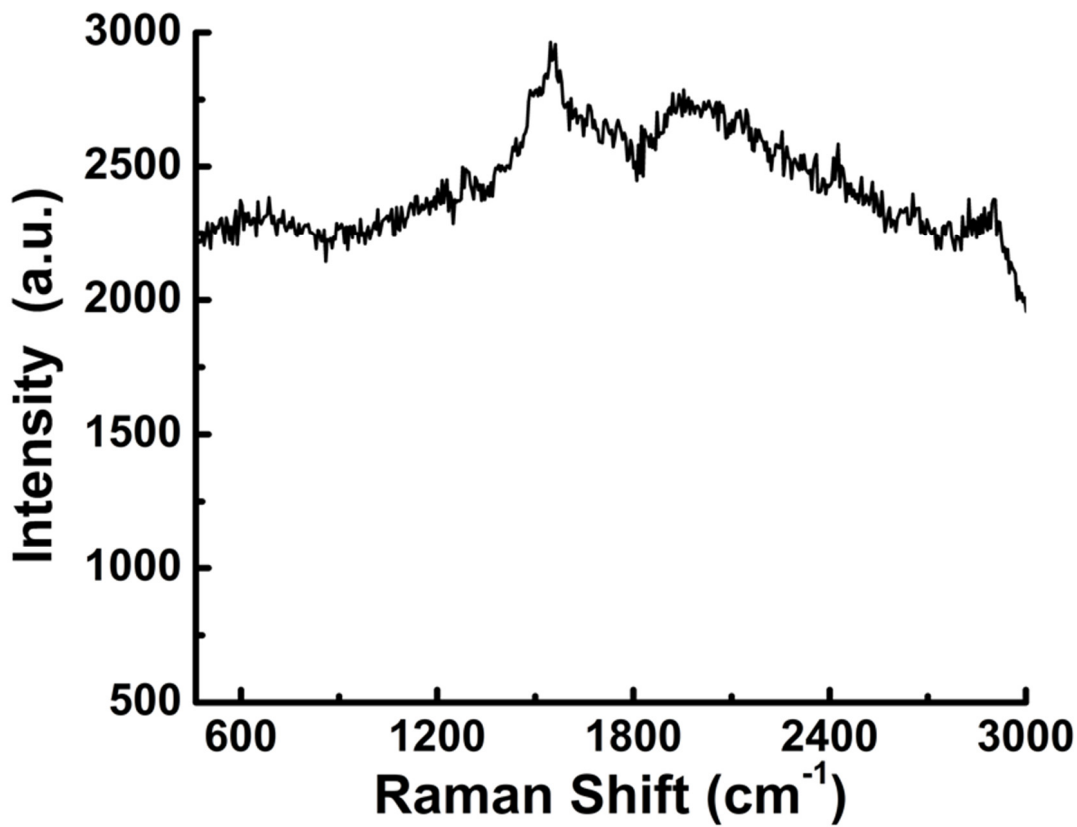

**Figure S5.** Inelastic emission from polyvinylpyrrolidone (PVP) coating on nanowire.

Figure S5 shows the spectrum collected from the end of AgNW placed on a gold mirror. One end of the nanowire was excited with 633 nm laser with polarization along the length of AgNW. The spectrum shows an inelastic background from PVP coating on the AgNW which is sandwiched between nanowire and mirror. The PVP coating is also present in the NW-NP cavity which also out-couples from the junction along with the SERS emission from the molecules.

## S6: Directionality calculation with different $\delta_1$ and $\delta_2$

| S. No. | $\delta_1$ (deg) | $\delta_2$ (deg) | Directionality (db) |
|---|---|---|---|
| 1 | 10 | 10 | 7.7 |
| 2 | 10 | 5 | 7.9 |
| 3 | 5 | 10 | 8.2 |
| 4 | 5 | 5 | 8.1 |

**Table S6.** Variation of directionality with a change in $\delta_1$ and $\delta_2$.

Table S6 shows the variation in the calculated directionality with a change in the value of $\delta_1$ and $\delta_2$ used in equation 1. The change in the calculated directionality is very less with a change in the values of $\delta_1$ and $\delta_2$.

## S7: Calculated Fourier plane image of emission from nanoparticle on mirror

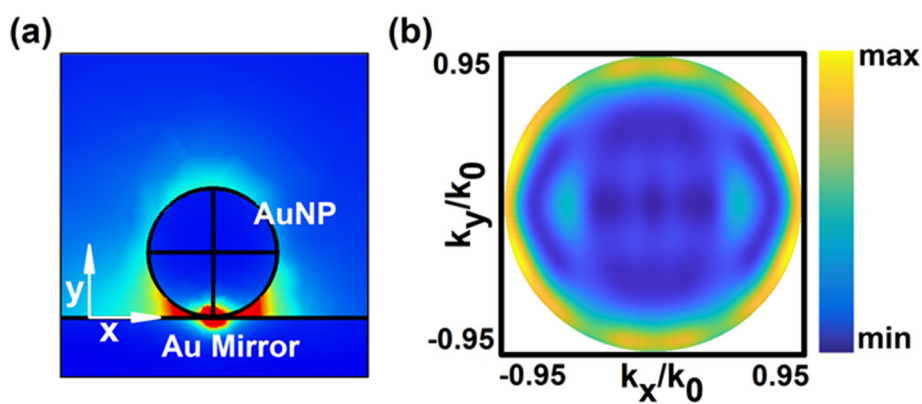

**Figure S7.** Calculated Fourier plane image of emission from nanoparticle on mirror. (a) Schematic of a nanoparticle on a mirror system. Oscillating x, y and z oriented dipoles at a wavelength of 703 nm are placed in the nanoparticle on mirror cavity and near-field electric field is calculated. (b) Fourier plane imaging of emission after incoherently adding the far-field signatures of individual dipoles. The emission is broad in nature and covers a large range of angles. The emission pattern is different from the experimentally obtained pattern shown in Figure S4 (f). The possible reason for this difference is that along with SERS from the molecules coated on the nanoparticle, there is also possibility of inelastic emission from the metal substrate or the particles which can out-couple at all the angles.

**References**

1. Y. Sun, B. Mayers, T. Herricks and Y. Xia, Nano letters, 2003, 3, 955-960.
2. P. Li, X. Yan, F. Zhou, X. Tang, L. Yang and J. Liu, Journal of Materials Chemistry C, 2017, 5, 3229-3237.
3. J. A. Kurvits, M. Jiang and R. Zia, JOSA A, 2015, 32, 2082-2092.
4. C. M. Dodson, J. A. Kurvits, D. Li and R. Zia, Optics letters, 2014, 39, 3927-3930.
5. A. B. Vasista, D. K. Sharma and G. P. Kumar, digital Encyclopedia of Applied Physics, 2018, 1-14.